\documentclass[conference]{IEEEtran}
\IEEEoverridecommandlockouts
\usepackage{cite}
\usepackage{amsmath,amssymb,amsfonts}
\usepackage{algorithmic}
\usepackage{graphicx}
\usepackage{textcomp}
\usepackage{comment}
\usepackage{fixltx2e}
\usepackage{ragged2e}
\usepackage{subcaption}
\usepackage{gensymb}
\usepackage{hyperref}
\usepackage{xcolor}
\usepackage{soul}
\usepackage[acronym]{glossaries}
\newacronym{cnn}{CNN}{Convolutional Neural Network}
\newacronym{map}{mAP}{mean Average Precision}
\newacronym{fmcw}{FMCW}{Frequency-Modulated Continuous-Wave}
\newacronym{cfar}{CFAR}{Constant False Alarm Rate}
\newacronym{rcs}{RCS}{Radar Cross Section}
\newacronym{dnn}{DNN}{Deep Neural Network}
\newacronym{ml}{ML}{Machine Learning}
\newacronym{stft}{STFT}{Short Time Fourier Transformation}
\newacronym{rd}{RD}{Range-Doppler}
\newacronym{roi}{ROI}{Region-of-Interest}
\newacronym{yolo}{YOLO}{You Only Look Once}
\newacronym{ra}{RA}{Range-Angle}
\newacronym{ad}{AD}{Angle-Doppler}
\newacronym{rad}{RAD}{Range-Angle-Doppler}
\newacronym{nn}{NN}{Neural Network}
\newacronym{bb}{BB}{Bounding Box}
\newacronym{dnms}{DNMS}{Distance-based non-max Suppression}
\newacronym{nms}{NMS}{non-max Suppression}
\newacronym{iou}{IOU}{Intersection over Union}
\newacronym{ap}{AP}{Average Precision}
\newacronym{mse}{MSE}{Mean Squared Error}
\newacronym{vru}{VRU}{Vulnerable Road User}
\newacronym{bev}{BEV}{Bird's-eye view}
\newacronym{cdc}{CDC}{Convolution-deconvolution}
\newacronym{fft}{FFT}{Fast Fourier Transform}
\newacronym{soc}{SOC}{System on Chip}

\def\BibTeX{{\rm B\kern-.05em{\sc i\kern-.025em b}\kern-.08em
    T\kern-.1667em\lower.7ex\hbox{E}\kern-.125emX}}

\newcommand{\Ali}{\textcolor{black}}
\begin{document}

\title{Raw Radar data based Object Detection and Heading estimation using Cross Attention

\thanks{$^{1}$The author is with the AVL GmbH, 93059 Regensburg, Germany
        {\tt\small\ ravi.kothari@avl.com}}
\thanks{$^{2}$The author is with Robert-Bosch GmbH, 31139 Hildesheim, Germany
        {\tt\small\ ali.kariminezhad@de.bosch.com}}
\thanks{$^{3}$The author is with the e:fs TechHub GmbH, 85080 Gaimersheim, Germany
        {\tt\small \ christian5.mayr@efs-auto.de}}
\thanks{$^{4}$The author is with Institute of Automatic Control, RWTH Aachen University, 52074 Aachen, Germany
{\tt\small\ h.zhang@irt.rwth-aachen.de}}
}
\author{\IEEEauthorblockN{Ravi Kothari$^{1}$, Ali Kariminezhad$^{2}$, Christian Mayr$^{3}$, Haoming Zhang$^{4}$}
}

\maketitle



\begin{abstract}

\Ali{Radar is an inevitable part of the perception sensor set for autonomous driving functions. It plays a gap-filling role to complement the shortcomings of other sensors in diverse scenarios and weather conditions.}  

In this paper, we propose a \gls{dnn} based end-to-end object detection and heading estimation framework using raw radar data. \Ali{To this end, we approach the problem in both a "Data-centric" and "model-centric" manner. We refine the publicly available CARRADA \cite{CarradaD} dataset and introduce Bivariate norm annotations. Besides, the baseline model is improved by a transformer \cite{AttenAll} inspired cross-attention fusion and further center-offset maps are added to reduce localisation error.} Our proposed model improves the detection~\gls{map} by 5\%, \Ali{while reducing the model complexity by almost 23\%}. For comprehensive scene understanding purposes, we extend our model for heading estimation. The improved ground truth and proposed model is available at \href{https://github.com/ravikothari510/crossattention_radar_detector}{Github}.

\end{abstract}

\begin{IEEEkeywords}
DNN, cross-attention, raw radar data, object detection, heading forecasting
\end{IEEEkeywords}

\section{Introduction}

\Ali{As the driving assistant functions mature, challenging requirements are expected to be fulfilled in the environment perception module. This means robust perception is anticipated in extreme weather conditions, urban environments, etc. This robustness requirement should be accompanied by a boosted perception performance while keeping the latency sufficiently low. To achieve these demanding goals, the integration of radar systems in the perception sensor set is indispensable.} \Ali{Current~\gls{fmcw} automotive radar systems deliver detections in the form of a sparse point cloud. These data points are filtered from the raw spectrum by the known~\gls{cfar} \cite{CFAR} algorithm. This intensity-based filtering process does not capture the complete object features and is solely triggered by their~\gls{rcs} and surrounding noise attributes. That means, the reflections from the obstacles with sufficiently high \gls{rcs} are likely to be detected as objects and the rest are considered as noise.} To capitalise on rich radar signature and extend its use for small objects detection and classification, recently many researchers have explored machine learning algorithms to directly process \gls{fft} maps.

Inspired by the computer vision community, researchers from radar society have also employed~\gls{cnn}-based models for detection, object classification and path prediction \cite{RadarID,RadarClass,TwoDCar,rrpn}. Nevertheless, the past works have been limited to individual aspects of radar properties, such as exploiting micro-Doppler for classification or utilizing only range-angle maps for object detection. This boundation is due to the difficulty in fusing distinct domains such as  Doppler channel with range-angle. On top of that, a competent radar-based perception module should also provide the object heading for effective scene understanding. To exploit the full potential of radar, we have incorporated the radar spectrum, i.e., raw data, in an end to end object detection and heading forecasting model. In this work, we consider the neural network proposed in RAMP-CNN \cite{Ramp} as our baseline for fusing all three radar channels. This baseline is further improved by the proposed cross-attention mechanism and center-offset integration. Due to limited angular resolution in radar, regressing tight 2D/3D bounding boxes, especially for \gls{vru} is infeasible. Thus for our purpose, a detection is defined as estimating an object center, similar to \cite{rodnet}. Furthermore, we improve the ground-truth representation using the Bivariate norm for the open-source dataset CARRADA \cite{CarradaD}. To improve upon the limited position accuracy of the baseline, we have incorporated a center-offset loss function. Moreover, for better scene understanding, a heading estimation is inferred with the same detector head.   
In summary, our main contributions are the following:
\begin{enumerate}
    \item Cross attention based Doppler feature fusion 
    \item Novel ground truth representation for object detection
    \item center-offset loss, to reduce distance error 
    \item Detector with heading estimation framework
\end{enumerate}

The paper is organised as follows, in section \ref{sec2} the latest results and the state-of-the-art works are presented. Followed by section \ref{sec3}, where we discuss the chosen dataset and its subsequent processing. In section \ref{sec4} the network architecture is explained and in section \ref{sec5} the proposed model is compared with the baseline. Finally, section \ref{sec6} is dedicated to conclusion and future scope of~\gls{dnn}-based radar detection. 

\section{Related Works}
\label{sec2}
Previous works \cite{sig_stft,rodnet,CarradaP,Ramp,YOLO_r,RAD,Bbox_r} on data-driven radar detection demonstrates the utility of~\gls{ml} in bringing out the full potential of radar sensor. These works generally focus on individual aspects of radar signals and process them with specific~\gls{nn} models. The authors in \cite{STFT} use~\gls{stft} heat maps as the input to the~\gls{cnn} to classify human activities. They have utilized the fact that dynamic objects have unique Doppler signatures, which could be captured by~\gls{cnn}. Furthermore, the authors in \cite{sig_stft} use a different form of micro-Doppler representation,~\gls{rd} maps as the input. Their model was a classifier with 3 classes (pedestrian, cyclist and car). These were initial methods to demonstrate the importance of micro-Doppler signatures in object classification. 

The authors in \cite{3DRadar} have shown a~\gls{dnn}-based radar detection using a two-stage network, where a~\gls{roi} is predicted and further processed through a classifier. It's an attempt at~\gls{dnn}-based semantic segmentation of a 3D point cloud.

Emulating the single-stage anchor-based object detectors (inspired by image processing), the authors of \cite{YOLO_r} have applied~\gls{yolo} \cite{yolo} on~\gls{rd} maps. On the same note \cite{Bbox_r} have used a ResNet \cite{resnet} based detector on~\gls{ra} maps. Although these works report improved detection capabilities, they are restricted in their approach by using individual radar features as either Doppler or \gls{ra} map.  

The authors in \cite{RAD} made the first attempt at rationally fusing different feature encodings. Their proposed model uses an autoencoder, which is capable of detecting vehicles in 2D \gls{bb}s. This method fuses the three two-dimensional \gls{ra}, \gls{ad} and \gls{rd} encoded feature maps by imposing redundancy in the missing dimension to bring all three feature maps in common space.

Due to low angular resolution in automotive radar, the accurate 2D \gls{bb} annotation of the radar spectrum is challenging. To resolve the problem of radar box annotation, the authors in \cite{rodnet} introduced a probabilistic map, which uses the target peaks to create Gaussian labels. This approach leads to a stable training process and circumvents the need for box annotation~\cite{centernet}. The authors in \cite{Ramp} have further extended Gaussian maps output to utilize all three radar channels. However, the adoption of an additional skip loss function for better utilization of Doppler branch, along with encoder feature replication similar to \cite{RAD} indicates this network architecture's limitations in efficiently extracting Doppler features. 

Heading estimation is critical in scene understanding, especially in urban scenarios. In the past \cite{TwoDCar} have shown that orientation estimation of a car is possible with radar, albeit its based on a point cloud dataset and evaluated on car class. We believe an object's doppler signature also contains its heading information i.e, velocity vector and it could be estimated even for small objects such as \gls{vru}. In this paper, we extend the detection network with further heading prediction. 


\section{Dataset}
\label{sec3}
On one hand, radar points are not dense enough to manually annotate 3D \gls{bb}. On the other, reading radar \gls{fft} spectrum's requires expertise. To overcome this hurdle, dataset benchmarks such as nuScenes \cite{nuscenes}, Oxford robocar \cite{oxf} and Astyx \cite{astyx}, which are all based on 3D sparse point clouds use cross-annotation with the help of camera and lidar sensors. The dearth of raw radar data could be attributed to the absence of off the shelf sensors and the vast size of data to be stored and hosted. Only in the last decade or so, with the advent for \gls{soc} and high bandwidth storage, it's now possible for researchers to record and share raw radar data reasonably.  

\Ali{CRUW \cite{rodnet} is one of the first datasets delivering annotated~\gls{ra} radar spectrum, where the labels are derived from image-based object detection and supervision of the accompanying camera. Improving on CRUW, the CARRADA dataset covers annotating the complete radar raw data (range, angle and Doppler), where the labels are provided as \gls{bb}s, sparse points and dense masks. The objects in the scenarios consist of multiple pedestrians, cyclists and cars interacting at different velocities, angles and ranges (5m -50 m).} On top of that the data includes \gls{vru} located in a noisy environment (snow). These scenarios are critical in highlighting the importance of radar where most of the camera and lidar sensor fails. CARRADA has also been proposed for temporal predictions~\cite{CarradaP}, which conforms with our goal for object heading estimation. 

CARRADA dataset labels are machine-generated (automatically labelled). It applies object detection (Mask R-CNN \cite{maskrcnn}) on camera images and further projects on \gls{bev} plane to get the annotation for radar raw data. Since the effectiveness of ML model is dependent on the quality of input data, we cannot directly rely on automatic labels. We pre-process this data to improve the annotation quality, create additional labels for heading estimation and transform 2D \gls{bb} to Gaussian maps.

\subsection{Dataset Pre-Processing}

Auto-labelling with a 2D image detector struggles with small objects, e.g., pedestrians at long distances. Such kinds of Mis-annotated ground truth lead to a sub-optimal network. In a data-centric manner, we first manually cross-check all the 7000 annotated frames and add additional 2D \gls{bb} for missed instances. These annotations are further projected on the \gls{ra} and \gls{rd} planes. In the end, our improved annotation had a 15\% more \gls{vru} as compared to the original dataset.  

Since, we set our goal not only for object detection, but also heading estimation, we had to supplement the CARRADA dataset with additional annotation. The heading ground truth is computed by finding the object's position in multiple frames, these points are further joined via a spline and the heading is computed as the differentiation of the curve at each instance.

\subsection{Bivariate Norm}

Replacing the 2D \gls{bb}s with a 2D Gaussian norms of unique variance has been found to improve the performance of radar based detector to some extent~\cite{rodnet}. However, applying a common Gaussian label for all the dimensions (range, angle, Doppler) and all the classes is far from optimal. This is due to the fact that, the radar signature changes with \gls{rcs} and object location. Hence, an ideal ground truth map should be flexible to reflect the real radar signatures. This guarantees that the optimizer will punish the network as per the signal strength. To achieve this, RODNet \cite{rodnet} tweaked the respective Gaussian annotations using multiple hyper-parameters that control the variance based on range, angle and object class. This method improves the ground truth, but it comes with an extra overhead of finding these hyper-parameters. 

To make an adaptive groundtruth, we propose the mapping of object-relevant raw spectrum with Gaussian maps. These maps required the Gaussian second-order moments. These moments are derived from the spectrum bounded with the proposed \gls{bb}. This process guarantees that the ground truth accurately reflects the original signal without using any extra parameters. This can be interpreted as a combination of semantic segmentation and Gaussian representation. Below are the steps showing the Bivariate norm mapping process:
\begin{enumerate}
    \item Normalise the \gls{bb} map by its max bin, i.e., in range 0-1.
    \item Noise separation by masking with a threshold.
    \item Calculate the covariance and center of the resultant masked bins.
    \item Project the Bivariate norm as the ground truth using the derived covariance and mean values.
\end{enumerate}

Given the range and angle first-order moments, $\mu_{r}$ and $\mu_{a}$ as object location and the derived range-angle covariance matrix (second-order moments) $\pmb{\Sigma}$, a bivariate Gaussian map is formulated as:
\begin{multline}
 f(r,a) = \frac{1}{2\pi \sigma_{r} \sigma_{a} \sqrt{1-\rho^2}} \cdot \exp \Biggl( \Biggl. - \frac{1}{2(1-\rho^2)} \biggl[ {\Bigl( \frac{r-\mu_{r}}{\sigma_{r}} \Bigr) }^2 \\
 -2\rho \Bigl( \frac{r-\mu_{r}}{\sigma_{r}} \Bigr) \Bigl( \frac{a-\mu_{a}}{\sigma_{a}} \Bigr) + {\Bigl( \frac{a-\mu_{a}}{\sigma_{a}}\Bigr) }^2 \biggr] \Biggr. \Biggr)
\end{multline} 

\begin{figure}[h]
    \centering
    {\includegraphics[width=1\linewidth]{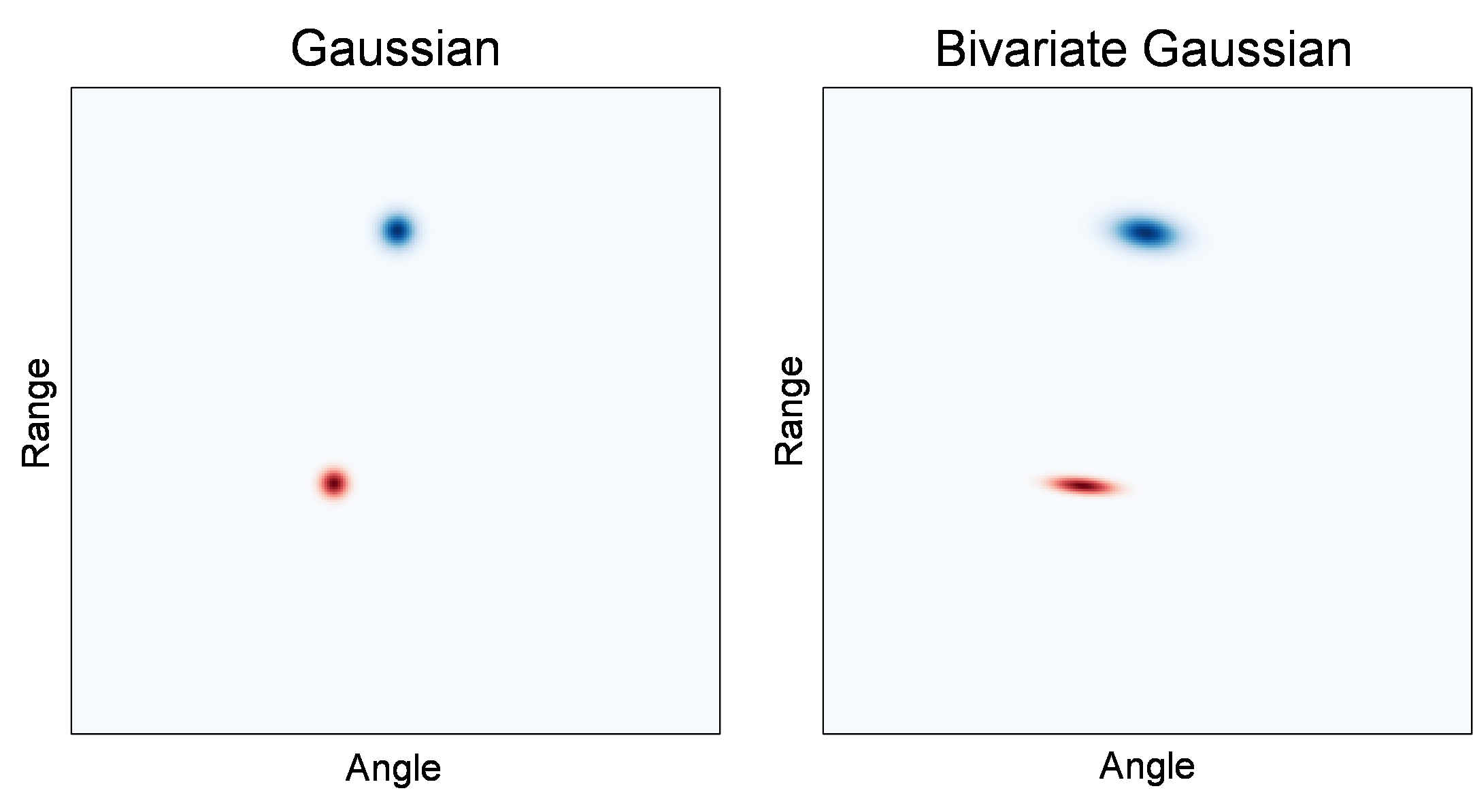}}
    \caption{uniform Gaussian vs adaptive variance Gaussian (Bivariate Norm) of a car (red) and pedestrian (blue)}
    \label{fig:bivargauss}
\end{figure}

Fig \ref{fig:bivargauss}, shows the variance directly captured from raw data. The angle axis has a high variance, indicating low angular resolution of the radar sensor. Similarly, with small objects (Pedestrians) at large distances, the range variance also increases. Hence, the proposed adaptive variance norm is a closer reflection of ground truth as compared to a hard-coded Gaussian patch. 

For the baseline network, we have considered correlation $\rho$ to be zero and the range and angle variances are assumed to be equal. 


\section{Network Architecture}
\label{sec4}

We aim to detect objects in~\gls{ra} coordinates, thus the network output will be of a similar shape as the input~\gls{ra} map. Most of the past works in input-output mapping have been inspired by medical imaging (semantic segmentation), denoising networks or some form of \gls{bb} detectors. Unlike image-based object detectors, where the network needs to go deep for enlarging the receptive fields. A radar signature remains fairly localised in \gls{ra} plane, thus even a network with small receptive field could perform well. 


\gls{cdc} type backbone belong to a family of autoencoders and are aimed at providing the degrees of freedom for spatial decoding of the encoded features in the same input dimension. The authors in \cite{Bbox} utilised a \gls{cdc} network for \gls{bb} detection. Correspondingly, RODNet \cite{rodnet} applied this approach with radar-based detection on the CRUW dataset, which delivered a 78\% average recall. Owing to \gls{cdc}'s light architecture and its adaptability to different task we have chosen CDC backbone for our model. 

\subsection{Fusion}

The incompatibility of Doppler and range-angle have driven the past fusion approaches to directly project Doppler intensities to RA maps \cite{CarradaP} or use 2D Doppler feature replication \cite{RAD, Ramp} on a RA axis before passing it through decoder. The former makes the network more reliant on RA maps and the latter creates a redundant fusion features. Hence, these models struggle with small targets, which inherits a consistent Doppler pattern but have low~\gls{rcs}. To realise the full potential of Doppler signature, we introduce a transformer \cite{AttenAll} based cross attention module, it uses Doppler features to attend on encoded RA maps.

Given a generalised feature map of size $ F\in R ^ {C\times H\times W}$, $F_{RA}, F_{RD}$ and $F_{AD}$ are the encoded features for cross attention block. The intermediate feature map $F^{\prime }$ is a matrix multiple of $F_{RD}$ and transposed $F_{AD}$, as shown in eq. \ref{inter}.
\begin{equation}
\label{inter}
F^\prime = F_{RD} \times F^T_{AD}
\end{equation}
Where $\times$ denotes matrix multiplication. $F^{\prime }$ is further normalised with softmax and a Hadamard (element wise) product  $\odot$ is applied with ${F_{RA}}$ feature map. 
\begin{equation}
    F^{\prime \prime} = softmax(F^\prime)\odot F_{RA}
\end{equation}
\begin{equation}
    F_{CA} = \|F^{\prime \prime} + F_{RA}\|_{layernorm}
\end{equation}

The output of a cross attention block $F_{CA}$ consists of a skip connection from the original $F_{RA}$ feature map which is added to $F^{\prime \prime}$. The resultant is normalised over layers, similar to the original attention block implementation in \cite{AttenAll}. This skip connection helps network, with stationary objects, where Doppler information is absent.

\subsection{Metrics}

\Ali{The network output, i.e., Gaussian maps from the decoder are passed through a peak detector.} If the detected maximum lies on the center of the kernel and has a confidence score greater than a pre-defined threshold, e.g., 0.1, it's considered a peak. Finally, these peak coordinates and intensity values are passed through~\gls{dnms} algorithm to remove overlapping detections. This approach is similar to~\gls{nms} from image-based detectors, here we switch the~\gls{iou} values with distance among peak pairs. In this work, a prediction is considered a true positive if it lies within a distance threshold of the ground truth. We evaluate our models with (2m and 1m) distance threshold. As compared to radar counter-parts such as more accurate Lidar 3D \gls{bb}, our association distance is limited due to low angular resolution and weak radar signatures of \gls{vru} at large distances. 

\subsection{Center-offset}

Owning to the fact that the network output is a rough estimate of Gaussian, the max kernel was sometimes detecting peaks away from the Gaussian center, this in turn lead to reduction in localization accuracy. To assist the network in finding Gauss center, we introduced a center-offset mask. A mask of size ($256 \times 256$)  for range and angle axis is created individually, and the normalised bin offset values are filled in a region of $9 \times 9$ around the Gauss center. The network is trained to regress this offset patch and during inference, all the detected peak are first shifted using their respective range-angle offset maps and then~\gls{dnms} is applied.

\begin{figure*}[!h]
    \centering
    {\includegraphics[width=1\textwidth]{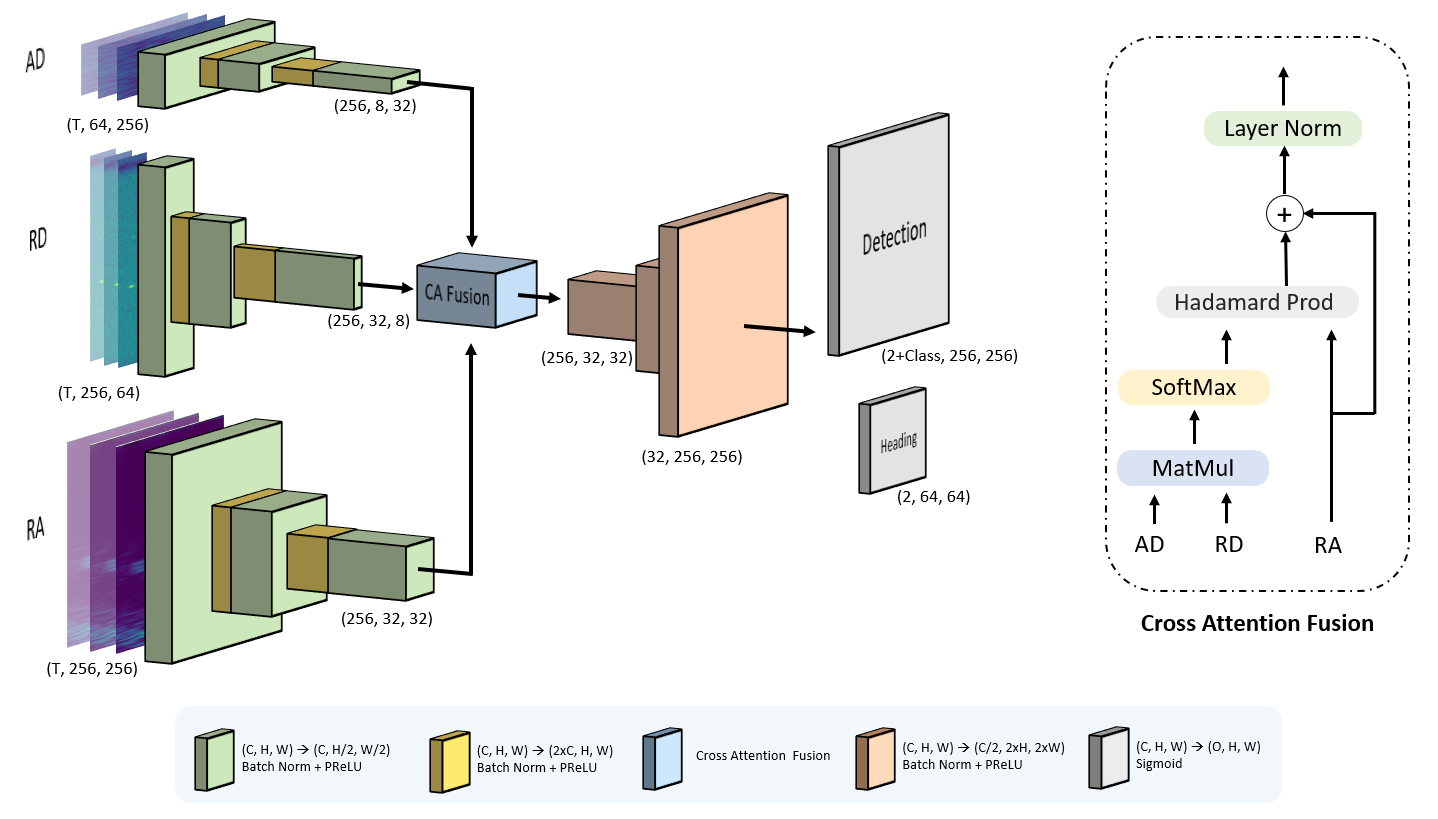}}
    \caption{Bivariate Cross Attention Network. Object detection purpose (T=1) and for heading estimation ablation study (T=1/3/5)}
    \label{fig:bivar_cross}
\end{figure*} 

\subsection{Heading}
\Ali{Radar range-angle-Doppler (RAD) signatures are dependent on the objects' position, orientation and velocity. Thus, theoretically based on these distinct fingerprints, it's possible to trace those patterns to determine objects' heading with single instance. In this paper, we adopt this observations and propose estimating objects' headings with a single RAD frame.} The heading angle ground truth is embedded in $\sin(\theta)$ and $\cos(\theta)$ maps. Since both the angles are complementary, together they can cover the complete range of ($-\pi,+\pi$). As the localisation ground truth is in the form of a Gaussian, similar representation is not feasible in heading annotation. Thus, we set the heading maps in a $64 \times 64$ map, considering that the network should detect the gauss center within an accuracy of 4 bins in the \gls{ra} map. The single-frame results are also further compared with multi-frames for analysing the impact of past frames in heading prediction.

\subsection{Architecture}

The proposed network consists of 6 encoder down-sampling layers and 4 layers of transpose convolutions for the decoder. Each layer consists of PReLU \cite{PRelu} activation and batch normalization \cite{BNorm}. The network is further boosted by the cross attention based fusion. Furthermore, the decoder has multiple heads for Bivariate norm, center-offset, and heading maps. Fig.\ref{fig:bivar_cross} illustrates the network architecture with the corresponding feature dimensions. As our work is targeting radar application in the automotive field, it's crucial to have a low computation and memory footprint. Thus, we develop and analyse the work with a single radar frame. As an exception, only the ablation study of heading estimation is conducted with multiple frames. The encoder's first layer is adapted to use stacked past frames (represented by T).

\subsection{Loss Function}

We train Bivariate norm mask with focal-loss \cite{floss} $L_{b}$, similarly center-offset $L_{c}$ uses focal-loss, but with masking over the patch region. Furthermore, the heading loss $L_{h}$ is determined to be~\gls{mse}. Total loss incorporates weighted individual losses. 

\begin{equation}
\label{eq:lossweight}
    L = w_{1}L_{b} + w_{2}L_{c} + w_{3}L_{h}
\end{equation}


\section{Evaluation}
\label{sec5}
In this section, we present the evaluation results and its analysis. Furthermore, the baseline is compared with the proposed network. After dataset pre-processing, almost 9000 instances are delivered. These instances are further split into train, validate and test sets (70\%, 10\%, 20\%).

\subsection{Baseline}

This work aims to showcase the importance of Doppler channels and introduce methods of effectively fusing doppler features and study the impact of adaptive ground truth representation via Bivariate norms.
The single \gls{ra} map baseline model is chosen as RODNet-CDC \cite{rodnet}, and for a \gls{rad} based method we have selected RAMP-CNN \cite{Ramp}. The term baseline model in this work refers to only the use of model architecture i.e, backbone + header. The original work consists of many intricacies such as the use of temporal inception layer in RODNet \cite{rodnet} or data prepossessing and all-perspective loss function in RAMP \cite{Ramp}, which is not under our purview. Thus our results do not reflect the competency of complete original methods but rather explore the inefficiency in fusion architectures and ground truth representation. Due to the absence of benchmark models in heading estimation with raw radar data, we evaluate our proposed model with different number of input RAD frames.

\subsection{Implementation}

PyTorch 1.9.0 framework is utilized for training and validation. All experiments were run on a Nvidia Quadro RTX 8000 GPU. To separate the model contributions from the pre-processing step, we first evaluate the baseline network using \textbf{cross attention} module with plain Gaussian labels and additionally center-offset and bivariate norms are introduced. The data augmentation (Gaussian noise, horizontal flip) is same for all the models. Batch size of 16 is used, with an initial learning rate of 1e-4, it's further reduced on plateau (factor=0.1, patience=4). Adam \cite{adam} is used as an optimizer. 80 epochs were found to be sufficient where the evaluation losses were achieving the minima.

\subsection{Detection Results}

\begin{table*}[t]
    \centering
    \label{tab:comparison}
    \begin{tabular}{|l|l|l|l|l|l|l|l|l|l|l|l|}
    \hline
        Network & Input & Ground truth & \multicolumn{8}{c|}{AP @ 2m and 1m} & Para. \\ \hline
         & & & \multicolumn{2}{c|}{Pedestrian} & \multicolumn{2}{c|}{Cyclist} & \multicolumn{2}{c|}{Car} & \multicolumn{2}{c|}{Overall} & \\ \hline
         RODNet-CDC \cite{rodnet} & RA & Gauss & 0.88 & 0.80 & 0.84 & 0.80 & 0.93 & 0.87 & 0.88 & 0.82 & 4.6 M \\ \hline
        RAMP-CNN \cite{Ramp} & RAD & Gauss & 0.89 & \textbf{0.82} & 0.86 & 0.79 & 0.93 & 0.88 & 0.89 & 0.83 & 14.23 M \\ \hline
        Cross Atten. & RAD & Gauss & \textbf{0.92} & 0.81 & 0.85 & 0.82 & 0.96 & 0.86 & 0.91 & 0.83 & 11.05 M\\ \hline
        Cross Atten. + offset loss & RAD & Gauss & 0.91 & 0.82 & 0.86 & 0.84 & 0.94 & 0.86 & 0.90 & 0.84 & 11.05 M\\ \hline
        Bivar Cross Atten. + offset loss & RAD & Bivariate & 0.91 & 0.8 & \textbf{0.93} & \textbf{0.88} & \textbf{0.97} & \textbf{0.91} & \textbf{0.94} & \textbf{0.86} & 11.05 M\\ \hline
    \end{tabular}
    \caption{Model performance and complexity comparison using single radar frame}
\end{table*}

\Ali{Cyclists and pedestrians are safety-critical classes hence, the term \gls{vru}, on top of that their detection is challenging due to their low \gls{rcs}. RODNet \cite{rodnet} delivers its highest~\gls{ap} (0.93) (corresponding to class car). This manifests to network dependence on high radar reflection for detection (vehicles poses high \gls{rcs} due to the metallic material and big size)}. The additional Doppler channel in RAMP-CNN \cite{Ramp} has improved the \gls{map} by 1\% over RODNet \cite{rodnet}, but this performance improvement comes with additional (3x) model parameters.

To isolate the contribution of Bivariate norm, we first show the cross attention network trained exclusively with plain Gaussian. Interestingly, our proposed cross attention module boosts the AP\textsubscript{2m} for classes car and pedestrian by \textbf{3\%} each. This shows the importance of the proposed cross attention in extracting Doppler features. But the same results are not reflected in the 1m distance threshold, this indicates the low accuracy detections in Gauss based representation.  Moreover, this module is realizable with a low complexity, which is granted due to applying a single decoder. The proposed fusion network results in a~\textbf{23\%} reduction in the number of parameters compared to RAMP-CNN. Introducing center-offset loss improves the \gls{ap} in short distance threshold (1m) but its performance reduces in 2m association. As the loss function is sensitive to the size of offset-patch, thus it might be further improved with an class dependent patch size. We understand the need for a bigger dataset and new \gls{dnn} models to improve pedestrian detection performance in 1m distance threshold. 

Combining the proposed "data-centric" approach of relabeling with Bivariate Gaussian and "model-centric" approach by integrating cross attention module and center-offset patch boosts the class "Cyclist" AP\textsubscript{2m} by 7\%. Compared to the baseline, the proposed network improves \gls{map} by \textbf{4\%}. The significance of center-offset maps could be depicted with the drop of \textbf{20\%} in \gls{mse} distance error (0.15m to 0.12m). The use of Doppler has also restricted the miss-classification rate below \textbf{1\%}, proving that the network doesn't exclusively rely on \gls{ra} maps signatures.

\subsection{Heading Estimation}

We consider the proposed network is deep enough to effectively extract heading features. This is achieved by adding an additional heading detector and adapting the encoder for accepting multiple frames. We report the heading accuracy in $\pm$45\degree,~$\pm$22.5\degree~and~$\pm$11.25\degree regions. These angular ranges are varied from coarse to fine, which could be further used in scene understanding with VRUs. Based on CARRADA~\cite{CarradaP} benchmark, it is observed that an average human stride could be captured in 5 radar frames. Hence, this is taken as the maximum number of input frames for the ablation study. 

\begin{table}[h]
    \centering
    \begin{tabular}{|l|l|l|l|l|}
    \hline
        No. of  & \multicolumn{3}{c|}{Heading Accuracy}\\\cline{2-4}
        input Frames & $\pm$45\degree & $\pm$22.5\degree  & $\pm$11.25\degree \\ \hline
        \hfil 1  & 71\% & 54\% & 35\%  \\ \hline
        \hfil 3  & 85\% & 67\% & 42\% \\ \hline
        \hfil 5  & \textbf{91\%} & \textbf{77\%} & \textbf{50\%}\\ \hline
    \end{tabular}
    \caption{Accuracy comparison of the network (Bivariate Cross Attention), while considering multiple past frames for heading estimation}
    \label{tab:orien_error}
\end{table}

\begin{figure}[h]
    \centering
    {\includegraphics[width=1\linewidth]{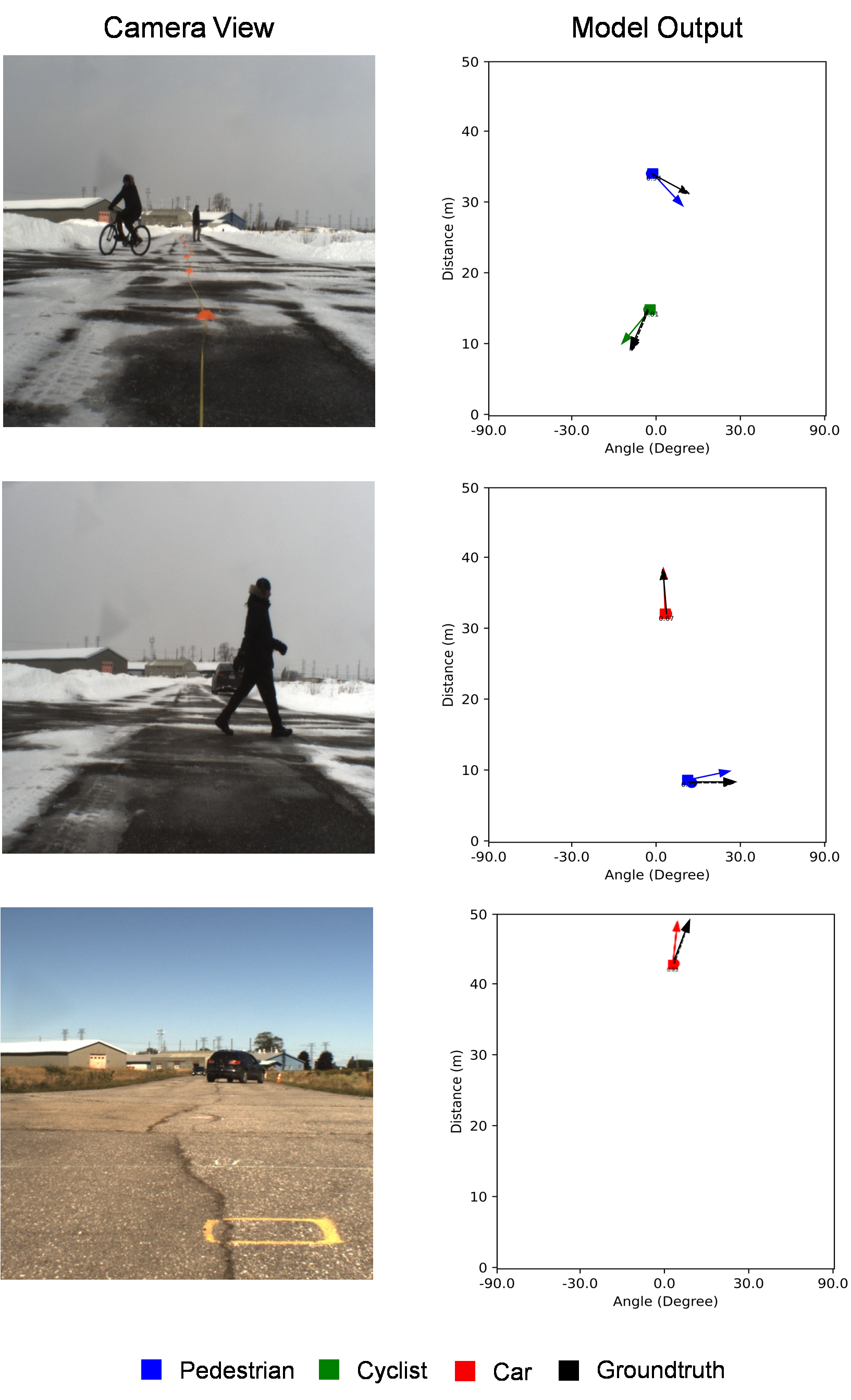}}
    \caption{Heading estimation and object detection using 5 past frames. The Black arrow indicates the ground truth heading and the colored arrow is the class specific predicted heading.}
    \label{fig:orien-ex}
\end{figure}

Table \ref{tab:orien_error} gives the absolute angular error between ground truth and predicted heading. As the number of frames increases the accuracy in fine region drastically goes up, which suggests the network is able to effectively fuse the maps in temporal space. Fig \ref{fig:orien-ex} shows the qualitative result of a 5 frame network. This example highlights the importance of radar in detecting occluded objects, where vision based sensors struggle. Furthermore, the network is capable of heading estimation of VRUs at long ranges without tracking. Proving our hypothesis, the single-frame network has also achieved an accuracy of 71\% for $\pm$45\degree. The low performance of the network in the fine heading region of $\pm$11.25\degree, suggests the coarseness of features embedded in \gls{rad} frame. 

\section{Conclusion}
\label{sec6}
\Ali{In this paper, we have proposed "data-centric" and "model-centric" novelties for object detection and heading estimation from radar raw data.} In this regard, we contributed to both data pre-processing and optimized network design. In the pre-processing phase, the objects are labelled via class-independent covariance Gaussian masks deduced from the dataset. For optimizing the network, an extra feature map known as center-offset feature map is exploited, which guides the network in finding Gaussian center. The combined model resulted in an overall \gls{map} improvement of 4\% compared to the baseline. This performance improvement comes along 23\% reduction in optimization parameters, i.e., network weights. Moreover, we have generalized the proposed network for object heading estimation, which eventually extracts the heading information from Range-Angle-Doppler maps. For future investigation, we would like to explore the performance gap in pedestrian detection and heading estimation in fine region. The next generations radar sensors with higher number of angle channels could be a way to resolve these issues. 

\bibliographystyle{IEEETran.bst}
\bibliography{IEEEabrv,reference}

\end{document}